\documentclass[prb,twocolumn]{revtex4}

\usepackage{graphics}
\usepackage{epstopdf}
\usepackage{amsmath}

\usepackage{soul}

\newcommand{\beq}{\begin{equation}}
\newcommand{\eeq}{\end{equation}}

\begin{document}

\title{Quantitative Wave-Particle Duality}
\author{Tabish Qureshi}
\email{tabish@ctp-jamia.res.in}
\affiliation{Centre for Theoretical Physics, Jamia Millia Islamia, New Delhi-110025, India.}

\begin{abstract}
The complementary  wave and particle character of quantum
objects (or quantons) was pointed out by Niels Bohr. This wave-particle
duality, in the context of the two-slit experiment, is now 
described not just as two extreme cases of wave and particle characteristics, but
in terms of quantitative measures of these natures. These
measures of wave and particle aspects are known to follow a duality
relation. A very simple and
intuitive derivation of a closely related duality relation is presented, which should
be understandable to the introductory student.
\end{abstract}

\maketitle

\section{Introduction}

The two slit experiment with particles is perhaps one of the most beautiful
experiments in physics. Particles going through a double-slit one by one,
accumulate to form an interference pattern.\cite{jonsson} With advancement
of technology, multislit diffraction has been demonstrated with large
molecules like $C_{60}$.\cite{buckyball}
Any quantum entity that shows the properties of both a particle and a wave,
is often called a quanton,\cite{bunge} and we will use this
nomenclature in this paper.
Although the quantons go
through the double-slit one at a time, the accumulated result of many
quantons shows an interference. It indicates that each quanton behaves
like a wave and interferes with itself. 

Objects like electrons are supposed to be
``indivisible" particles, unlike photons and phonons which one believes to be
actually waves, but occasionally localized into packets which look like
particles. Being ``indivisble" particles, one would naively imagine that these
quantons pass through one of the two slits, and not both. However, if
they yield an interference pattern, these particles must somehow be
interacting with both the slits at the same time, just like a wave. Trying
to resolve this issue, if one tries to experimentally find out which
of the two slits the quanton passed through, the interference disappears.
The quanton passing through one of the two slits, is associated with the
particle nature, whereas the interference seen in cumulative results
is associated with the wave nature. Wave and particle natures can only
be seen one at a time, never simultaneously. This was formalized by Niels
Bohr as the principle of complementarity.\cite{bohr} Einstein had tried
to argue against such a principle, and had proposed a thought experiment
which, he claimed, showed wave and particle natures in the same experiment.
\cite{tqeinstein} Bohr pointed out the flaw in Einstein's argument and
the principle of complementarity stood its ground.

It was recognized later that the wave and particles natures are not
mutually exclusive in the strict sense of the word. It is possible to
get partial information on which slit a quanton went through, and
get an interference pattern which is not sharp. In other words wave-particle
duality in Bohr's principle can be stated, not just in terms of mutual
exclusivity of purely particle and purely wave natures, but in terms of
quantitative measures of these natures.\cite{wootters,greenberger,vaidman,englert}
Wave-particle duality in terms of these quantitative measures of wave and
particle natures is now described by the following duality relation\cite{englert}
\begin{equation}
{\mathcal V}^2 + {\mathcal D}^2 \le 1,
\label{egy}
\end{equation}
where  ${\mathcal D}$ is a path distinguishability and ${\mathcal V}$ the
visibility of the interference pattern. Both these quantities vary between
0 and 1, and the above relation quantifies how visible will the interference
be if the two paths through the two slits can be distinguished with a
distinguishability ${\mathcal D}$. It should be emphasized here that
distinguishability is necessarily associated with a measuring apparatus
used to find out which of the two slits a quanton went through.
Distinguishability ${\mathcal D}$ being 1
would imply that one can say with absolute certainty which of the two slits
the quanton went through. The quanton going through a particular slit
implies that it behaves like a particle, in the sense described in the
preceding paragraph. No interference is seen in such a scenario, as the
interference visibility ${\mathcal V}$, as seen from (\ref{egy}), can only be 0.
On the other hand, if ${\mathcal V}$ is 1, the path distinguishability
${\mathcal D}$ can only be 0. In this situation one cannot tell which of
the two slits the quanton went through. The reader might wonder if this is
just the inability of the experimenter to tell which slit the quanton went
through, and if the quanton actually takes one path or the other. We would
like to assert here that it is incorrect to assume that the quanton always
takes one path or the other. Some prefer to assume that, left to itself,
the quanton will take both the paths, although quantum mechanics does not
say anything about it. Strictly speaking, one should assume the following.
Even if we have made a measurement on a hundered quantons, each of
which has told us which slit each of them passed through, we cannot
say if the next quanton went through only one of the two slits (irresepective of
which) without making a measurement. Unperformed experiments have no
results.\cite{unperformed}

Going further, one might wonder what partial distinguishability (${\mathcal D}$
having value between 0 and 1) means for a single quanton, and whether each
quanton contributes partially to interfrence. Or does the duality relation mean
that some quantons give full which-path information and do not contribute
to interference and some give zero which-path information and contribute
fully to interference? These are the questions a student is likely to
wonder, since the interference is built up by quantons going through
one by one. We would like to address this question and suggest a
picture to visualize. In the following, we analyze a thought modification of
the two-slit experiment with provision for path detection, and give a
novel derivation of a duality relation closely related to Eq. (\ref{egy}).

\section{Full which-path information}

It can be demonstrated quite simply that if there is a path-detector
which gains information about which of the two slits the quanton passed through,
the interference will be completely destroyed.\cite{eraser1, durr}
Suppose that the state of a quanton, passing through a
double-slit, is given by
\begin{equation}
             |\Psi\rangle = {1\over\sqrt{2}}(|\psi_1\rangle + |\psi_2\rangle),
\end{equation}
where $|\psi_1\rangle$ and $|\psi_2\rangle$ are the amplitudes of the quanton
passing through slit 1 and 2, respectively. The probability of finding the
quanton at a point x on the screen, is given by
\begin{eqnarray}
 |\langle x|\Psi\rangle|^2 &=& {1\over 2}(|\langle x|\psi_1\rangle|^2 + |\langle x|\psi_2\rangle|^2 \nonumber\\
&&+
               \langle \psi_1|x\rangle\langle x|\psi_2\rangle + \langle\psi_2|x\rangle\langle x|\psi_1\rangle).
\end{eqnarray}
In the above, $\langle x|\Psi\rangle$ is the wavefunction, also represented
by $\Psi(x)$.
The last two terms represent interference. Now, let us suppose that
we have a quantum path-detector also included in the setup. Without
going into the details of what such a path-detector might look like,
we just assume that a quanton going through slit 1 leaves the detector in
a state $|d_1\rangle$ and that going through slit 2 leaves the detector in
a state $|d_2\rangle$. If this detector is capable of detecting
which path the quanton went through, according to von Neumann's recipe
of a quantum measurement, the states of the detector should get correlated
with the two paths of the quanton.\cite{neumann} The state of the quanton
and the path-detector combined, will necessarily be entangled, and will be
of a form
\begin{equation}
|\Psi\rangle = {1\over\sqrt{2}}(|\psi_1\rangle|d_1\rangle + |\psi_2\rangle|d_2\rangle).
\label{corrstate}
\end{equation}
Measuring an observable which has two eigenstates $|d_1\rangle, |d_2\rangle$,
with different eigenvalues, will indicate that the quanton went through a
particular slit. 
A measurement of that observable yielding $|d_1\rangle$ will
lead to a definite conclusion that the quanton passed through slit 1,
and likewise for $|d_2\rangle$. This, however, does not imply that the
quanton goes through only one of the two slits, even when no path-detection
is made.
The probability of finding the quanton at a position $x$, is given by
\begin{eqnarray}
     |\langle x|\Psi\rangle|^2 &=& {1\over 2}(|\langle x|\psi_1\rangle|^2
\langle d_1|d_1\rangle + |\langle x|\psi_2\rangle|^2 \langle d_2|d_2\rangle\nonumber\\
&&+ \langle \psi_1|x\rangle\langle x|\psi_2\rangle\langle d_1|d_2\rangle + \langle\psi_2|x\rangle\langle x|\psi_1\rangle\langle d_2|d_1\rangle).\nonumber\\
\end{eqnarray}
The last two terms which would have given interference, are killed by the
orthogonality of $|d_1\rangle$ and $|d_2\rangle$, and one is reduced to
\begin{eqnarray}
     |\langle x|\Psi\rangle|^2 &=& {1\over 2}(|\langle x|\psi_1\rangle|^2 + |\langle x|\psi_2\rangle|^2).
\end{eqnarray}
An important point to
be noted here is that the actual
path measurement
is not even necessary here. The mere existence of which-way information,
or mere possibility of a path measurement, is enough to destroy
interference. 

The preceding discussion implies that finding the detector state (say) $|d_1\rangle$
indicates that the quanton went through slit 1. But one may wonder if
the quanton {\em really} went through slit 1, and if one can verify that.
The problem becomes more complicated if the quanton hits the screen first
and the path-detector states are looked at later. These are complex
interpretational issues, and a consensus on these has not been arrived at.
We would not delve into this issue here, and only refer the reader to a
debate on it.\cite{esw,mohrhoff}
For our purpose, we would continue to assume that finding a path-detector
state $|d_1\rangle$ would imply quanton going through the upper slit,
to the extent that is implied by the correlated state (\ref{corrstate}).

\section{Partial which-path information}

\subsection{Unambiguous quantum state discrimination}

Suppose now that the path-detector has states $|d_1\rangle$ and
$|d_2\rangle$ which are not orthogonal. In such a situation, no observable
exists for which $|d_1\rangle$ and $|d_2\rangle$ are eigenstates with 
different eigenvalues. Hence it is not possible to distinguish between 
the two states perfectly. Consequently they cannot be used to distinguish
between the two paths of the quanton.  As example, recently a two-slit interference experiment
was carried out at the molecular level where the atomic slits were
movable. Depending on which slit the particle passed through, the
atomic slits could recoil in two possible directions and thus carry the
information about the passing particle.\cite{liu}

In this case the role of $|d_1\rangle$ and $|d_2\rangle$ is played by
the momentum states of the counter-propagating atomic slits. If the two
momentum states are not distinct, or orthogonal, they cannot be used
to discern which slit the particle went through.

In the following we will describe a strategy which is the best one
to distinguish between two non-orthogonal states {\em without
error}. It goes by the name of unambiguous quantum state discrimination
(UQSD).\cite{uqsd,dieks,peres,jaeger2,bergou}

The state of the quanton plus the path-detector is still given by
(\ref{corrstate}), but now $|\langle d_1|d_2\rangle| \neq 0$.
For simplifying the analysis, we assume that $\langle d_1|d_2\rangle$ is real
and non-negative.
Since the detector design is in the experimenter's control, this can
always be arranged.
Let there
be a two-state {\em ancilla} system which interacts with the path-detector.
This interaction is characterized by the following properties,
\begin{eqnarray}
{\mathbf U_a} |d_1\rangle|a_0\rangle &=& \alpha|p_1\rangle|a_1\rangle
 + \beta |q\rangle|a_2\rangle \nonumber\\
\textrm{and}\nonumber\\
{\mathbf U_a} |d_2\rangle|a_0\rangle &=& \alpha|p_2\rangle|a_1\rangle
 + \beta |q\rangle|a_2\rangle ,
\end{eqnarray}
where the path-detector states $\langle p_1|p_2\rangle=0$, and the ancilla
states $\langle a_1|a_2\rangle=0$. 
It can be shown that as long as $\langle d_1|d_2\rangle$ is real and non-negative,
such an interaction always exists.\cite{uqsd,dieks,peres}
The two parameter
can be easily shown to be $|\beta|^2 = \langle d_1|d_2\rangle$
and $|\alpha|^2 = 1 - \langle d_1|d_2\rangle$.

Now, if in the measurement of the ancilla, one gets the state $|a_1\rangle$, the
corresponding path-detector states will be $|p_1\rangle$ or $|p_2\rangle$,
which are orthogonal. Measuring a suitable observable of the path-detector
will unambiguously tell us whether the state is $|p_1\rangle$ or $|p_2\rangle$,
and consequently, whether the original state was
$|d_1\rangle$ or $|d_2\rangle$. However, if the ancilla measurement yields
$|a_2\rangle$, the corresponding path-detector states
$|q\rangle$ are identical, and one cannot distinguish
between the two. In this case the process of distinguishing between 
$|d_1\rangle$ and $|d_2\rangle$ fails. If the states $|d_1\rangle$ and
$|d_2\rangle$ occur with probability $1/2$ each, the probability of failure
to distinguish is just $|\beta|^2 = \langle d_1|d_2\rangle$.

Hence the probability of successfully {\em unambiguously} distinguishing
between $|d_1\rangle$ and $|d_2\rangle$ is 
\begin{equation}
P = 1 - \langle d_1|d_2\rangle .
\end{equation}
The probability given by the above relation is the maximal probability with
which an apparatus interacting with the d-system can {\em unambiguously}
answer whether it is in state $|d_1\rangle$ or $|d_2\rangle$.\cite{peres}

\subsection{Distinguishing between the quanton paths}

Coming back to our problem of distinguishing between the two paths of the
quanton, let us start from the state Eq.~(\ref{corrstate}).
If, for the time being, one is only interested in the state of the
path-detector, one may ignore the states of the quanton. Thus, the reduced density matrix is
\begin{eqnarray}
\rho_r &=& \sum_{i=1}^{2} \langle\psi_i| |\Psi\rangle\langle\Psi| |\psi_i\rangle
\nonumber\\
  &=& {1\over 2} |d_1\rangle\langle d_1| + {1\over 2} |d_2\rangle\langle d_2|,
\end{eqnarray}
where we have used the orthogonality of $|\psi_1\rangle$ and $|\psi_2\rangle$.
The above indicates that the path-detector, for all practical purposes, is
in a mixed state, where it may be thought to be randomly occuring be in state
$|d_1\rangle$ or $|d_2\rangle$ with equal probability, provided one can
distinguish between the two. The problem then reduces to
distinguishing whether one has $|d_1\rangle$ or $|d_2\rangle$, and UQSD is
suited to provide an answer in this situation.

There is a subtle assumption in this interpretation that the states
 $|d_1\rangle$ or $|d_2\rangle$ occur as pure states randomly, whereas they
actually occur in the entangled state Eq.~(\ref{corrstate}). This assumption may
be justified by the observation that if one is dealing with a 
entangled system,
by looking at just one part of the system, one cannot distinguish between
a mixed state and an entangled state.

We let the ancilla interact with
the path-detector, with the starting state being written as
\begin{equation}
|\Psi_i\rangle = {1\over\sqrt{2}}(|\psi_1\rangle|d_1\rangle + |\psi_2\rangle|d_2\rangle)
|a_0\rangle.
\label{corrstatei}
\end{equation}
The interaction operator, for the ancilla and the path-detector, acts on the
full entangled state, and the result is
\begin{eqnarray}
|\Psi_f\rangle &=& {\mathbf U_a}|\Psi_i\rangle\nonumber\\
&=& {1\over\sqrt{2}}{\mathbf U_a}(|\psi_1\rangle|d_1\rangle + |\psi_2\rangle|d_2\rangle)
|a_0\rangle \nonumber\\
&=& {1\over\sqrt{2}}|\psi_1\rangle(\alpha|p_1\rangle|a_1\rangle
 + \beta |q\rangle|a_2\rangle) \nonumber\\
&&+ {1\over\sqrt{2}}|\psi_2\rangle(\alpha|p_2\rangle|a_1\rangle
 + \beta |q\rangle|a_2\rangle )\nonumber\\
&=& {\sqrt{1-\langle d_1|d_2\rangle}\over\sqrt{2}} (|\psi_1\rangle|p_1\rangle
           + |\psi_2\rangle|p_2\rangle)|a_1\rangle \nonumber\\
&&+ 
 {\sqrt{\langle d_1|d_2\rangle}\over\sqrt{2}}
 (|\psi_1\rangle + |\psi_2\rangle)|q\rangle|a_2\rangle.
\label{interf}
\end{eqnarray}
First we make sure that introducing the ancilla system does not affect the
visibility of interference. The probability density of the quanton hitting
the screen at a position $x$ is given by
\begin{eqnarray}
|\langle x|\Psi_f\rangle|^2 &=& 
(1-\langle d_1|d_2\rangle){1\over 2}(|\langle x|\psi_1\rangle|^2
|\langle p_1|p_1\rangle|^2 \nonumber\\
&&+ |\langle x|\psi_2\rangle|^2
|\langle p_2|p_2\rangle|^2)\nonumber\\
&&+\langle d_1|d_2\rangle{1\over 2}(|\langle x|\psi_1\rangle|^2
+ |\langle x|\psi_2\rangle|^2 \nonumber\\
&&+ \langle x|\psi_1\rangle\langle\psi_2|x\rangle
+ \langle x|\psi_2\rangle\langle\psi_1|x\rangle) |\langle q|q\rangle|^2\nonumber\\
&=& {1\over 2}\left(|\langle x|\psi_1\rangle|^2 + |\langle x|\psi_2\rangle|^2\right.
\nonumber\\
&&\left.+\langle d_1|d_2\rangle\{\langle x|\psi_1\rangle\langle\psi_2|x\rangle
+ \langle x|\psi_2\rangle\langle\psi_1|x\rangle\}\right),\nonumber\\
\label{prob}
\end{eqnarray}
where we have used the fact that $|p_1\rangle, |p_2\rangle, |q\rangle$ are
normalized. Visibility of the interference fringes is conventionally defined
as ${\mathcal V} = {I_{max} - I_{min} \over I_{max} + I_{min} }$,
where $I_{max}$ and $I_{min}$ are the maximum and minimum values of
intensity in a close neighborhood.\cite{born} In a two-slit interference, the ideal
visibility can be worked out to be just the factor
multiplying the term $\psi_1^*(x)\psi_2(x)+\psi_2^*(x)\psi_1(x)$.\cite{tqeinstein} So the
visibility ${\mathcal V}$ can be just read off from Eq.~(\ref{prob}) as
$\langle d_1|d_2\rangle$. It is straightforward to check that if one starts
from Eq.~(\ref{corrstate}) instead of Eq.~(\ref{interf}), one obtains the same visibility.

The quantons hitting the screen can be divided into two sub-ensembles
according the ancilla states $|a_1\rangle$ and $|a_2\rangle$. Quantons
correlated with the ancilla state $|a_1\rangle$ are in the state
\begin{equation}
\langle a_1|\Psi_f\rangle = {1\over\sqrt{2}}(|\psi_1\rangle|p_1\rangle
           + |\psi_2\rangle|p_2\rangle).
\end{equation}
In this state the quanton path amplitudes are correlated with orthogonal
states of the path-detector. Hence these quantons will not show any
interference, as can be checked by evaluating
$|\langle a_1|\langle x|\Psi_f\rangle|^2$.
However, for each of these quantons, measuring an observable of the 
path-detector whose 
non-degenerate eigenstates are $|p_1\rangle, |p_2\rangle$, unambiguously
tells us which slit the quanton passed through.

Quantons correlated with the ancilla state $|a_2\rangle$ are in the state
\begin{equation}
\langle a_2|\Psi_f\rangle = {1\over\sqrt{2}} (|\psi_1\rangle + |\psi_2\rangle)|q_1\rangle,
\end{equation}
and the probability distribution of these quantons on the screen is given by
\begin{equation}
|\langle a_2|\langle x|\Psi_f\rangle|^2 = {1\over{2}} |\langle x|\psi_1\rangle + \langle x|\psi_2\rangle|^2.
\end{equation}
These quantons will show full interference.
All of this can also be verified in 
an experiment by correlating the quantons hitting the screen with the
measurement results of the ancilla.
There is a subtle point however, which should be mentioned here. Once the correlation
between the quanton and the ancilla has been established by the state Eq.~(\ref{interf}),
it does not matter if the ancilla is measured first and the quanton is
detected on the screen later or vice versa. Both cases will show exactly the same correlation in the measurement results. In fact, one may choose which
operator of the ancilla to measure, well after the quanton has been
registered on the screen. The state in Eq.~(\ref{interf}) ensures that in an
actual experiment, those correlations will be seen for sure. Such issues
have also been discussed earlier.\cite{esw,mohrhoff}
In this sense, given the state in Eq.~(\ref{interf}), one can talk about which states of the quantons are 
correlated with which states of the ancilla, without doing an actual
measurement.

\subsection{Quantitative wave-particle duality}

From Eq.~(\ref{interf}), one can see that the fraction of quantons which contribute
to interference, is $\langle d_1|d_2\rangle$. But the fraction of quantons
giving rise to interference should intuitively be the {\em visibility} of
the interference pattern. Indeed it is identical to the visibility obtained
in the preceding subsection. So the interference visibility, denoted by
${\mathcal V}$ is given by
\begin{equation}
{\mathcal V} = \langle d_1|d_2\rangle .
\label{V}
\end{equation}
The above relation for visibility agrees with the one derived by Englert.\cite{englert}

We can define path-distinguishability ${\mathcal D}_Q$ as the fraction of
quantons for whom one can unambiguously tell which slit they passed through,
{\em in the best case}.
For a single quanton, path-distinguishability ${\mathcal D}_Q$ is definied
as the {\em maximum} probability with which one can unambiguously tell which
slit it passed through. From Eq.~(\ref{interf}), that fraction is just
$1 - \langle d_1|d_2\rangle$. Hence path-distinguishability is given by
\begin{equation}
{\mathcal D}_Q = 1 - \langle d_1|d_2\rangle,
\label{D}
\end{equation}
and from Eq.~(\ref{V}) and (\ref{D}) one can write
\begin{equation}
{\mathcal D}_Q + {\mathcal V} = 1.
\end{equation}
The above result is a direct consequence of the fact that the fraction of
quantons which give rise to interference is $\langle d_1|d_2\rangle$,
and the fraction of quantons for which one can tell {\em for sure} which
slit they came through, is $1 - \langle d_1|d_2\rangle$.
It should be noted that these fractions can be inferred from Eq.~(\ref{interf}) without doing an
actual measurement on the ancilla. However, in order to find out for which
quantons one can get full which-path information, an actual
measurement on the ancilla must be performed.  Getting a state $|a_1\rangle$ of the ancilla would
imply that for the particular quanton, one can tell for sure which slit it
came from. If one gets a state $|a_2\rangle$, one cannot tell which slit
that particular quanton came from.

If real experimental factors are taken into account, both visibility and
path-distinguishability will have reduced values. Hence we can write the
following inequality
\begin{equation}
{\mathcal D}_Q + {\mathcal V} \le 1.
\label{duality}
\end{equation}
This is a duality relation which quantifies wave-particle duality,
or complementarity, in a two-slit interference experiment.

In terms of the path-detector states $|d_1\rangle$ and $|d_2\rangle$,
the distinguishability introduced by Englert has the form\cite{englert}
\begin{equation}
{\mathcal D} = \sqrt{ 1 - |\langle d_1|d_2\rangle|^2},
\end{equation}
and Englert's distinguishability ${\mathcal D}$ can be related to ${\mathcal D}_Q$
by the relation
\begin{equation}
{\mathcal D}_Q = 1 - \sqrt{1 - {\mathcal D}^2}
\end{equation}
Substituting the above in Eq.~(\ref{duality}) gives
\begin{equation}
{\mathcal V} \le \sqrt{1 - {\mathcal D}^2}.
\end{equation}
Squaring both side, one get
\begin{equation}
{\mathcal V}^2 + {\mathcal D}^2 \le 1,
\end{equation}
which is identical to Eq.~(\ref{egy}).
So, when the quantity ${\mathcal D}$ of Eq. (\ref{egy})
is evaluated for the case of pure detector states, the resulting relation
between $\langle d_1|d_2\rangle$ and ${\mathcal V}$ is the same as the
relation given by Eq. (\ref{duality}). In this sense Eqs. (\ref{egy})
and (\ref{duality}) are very closely analogous.

\section{Discussion}

We have discussed a thought modification of a two-slit interference 
experiment, where the which-path information of quantons is extracted
using UQSD. The analysis shows that all the quantons passing through the
double-slit, and registering on the screen, can be split into two sub-ensembles,
depending on the measurement results of the ancilla. For quantons falling in
the first sub-ensemble, one can unambiguously tell for each quanton which
slit it passed through. These quantons do not contribute to interference.

For the quantons falling in the second sub-ensemble, one cannot tell
which slit each of them passed through, but they all contribute to
interference. Interference visibility intuitively should be just the
fraction of quantons contributing fully to interference.
For simply calculating this fraction, without specifying which particular
quantons can give full which-path information, one need not even do an
actual measurement on the ancilla.

Using just the above arguments, we have derived a bound on the sum of
the path-distinguishability and interference visibility.
The derived duality relation expresses, in terms
of unambiguous quantum state determination, a wave-particle trade-off closely
analogous to a well-known duality relation.\cite{englert} 
UQSD can also
be done for more than two states. Using UQSD, the duality relation has
also been extended to interference experiments involving more
than two slits.\cite{3slit,nslit}

\end{document}